\documentclass[12pt,preprint]{aastex}

\newcommand{\hi}{H{\sc i}}
\newcommand{\htwo}{H$_2$~1-0S(1)}
\newcommand{\fe}{[Fe~{\sc ii}] 1.644~$\mu$m}
\newcommand{\brg}{Br$\gamma$}
\newcommand{\kms}{~km~s$^{-1}$}

\slugcomment{Not to appear in Nonlearned J., 45.}

\shorttitle{AO assisted spectroscopy of SVS\,13}
\shortauthors{Davis et al.}

\begin{document}


\title{Adaptive Optics Assisted Near-Infrared 
Spectroscopy of SVS\,13 and its Jet\altaffilmark{1}}


\author{C.J. Davis}
\affil{Joint Astronomy Centre, 660 N. A'oh\={o}k\={u} Place, Hilo, 
Hawaii 96720, USA}
\email{c.davis@jach.hawaii.edu}

\author{B. Nisini}
\affil{INAF-Osservatorio Astronomico di Roma, Via di Frascati 33, 
I-00040 Monteporzio Catone, Italy}

\author{M. Takami and T.-S. Pyo}
\affil{Subaru Telescope, 650 N. A'oh\={o}k\={u} Place, Hilo, 
Hawaii 96720, USA}

\author{M.D. Smith}
\affil{Armagh Observatory, College Hill, Armagh, N. Ireland}

\author{E. Whelan and T.P. Ray}
\affil{Dublin Inst. Advanced Studies, 5 Merrion Square, Dublin 2, Ireland} 

\author{A. Chrysostomou}
\affil{Centre for Astrophysics Research, University of Hertfordshire, 
Hatfield, Herts AL10 9AB, UK}


\altaffiltext{1}{Based on Observations collected at the European
Southern Observatory, Paranal, Chile (ESO programmes 074.C-0408).}


\begin{abstract} 

We present long-slit H- and K-band spectroscopy of the low-mass
outflow source \object{SVS 13}, obtained with the adaptive-optics
assisted imager-spectrometer {\em NACO} on the {\em VLT}.  With a
spatial resolution of $<$0.25\arcsec\ and a pixel scale of
0.027\arcsec\ we precisely establish the relative offsets of H$_2$,
[Fe~{\sc ii}], CO, H{\sc i} and Na{\sc i} components from the source
continuum.  The H$_2$ and [Fe {\sc ii}] peaks are clearly associated
with the jet, while the CO, H{\sc i} and Na{\sc i} peaks are spatially
unresolved and coincident with the source, as is expected for emission
associated with accretion processes.  The H$_2$ profile along the slit
is resolved into multiple components, which increase in size though
decrease in intensity with distance from the source.  This trend might
be consistent with thermal expansion of packets of gas ejected during
periods of increased accretion activity.  Indeed, for the brightest
component nearest the source, proper motion measurements indicate a
tangential velocity of 0.028\arcsec /year.  It therefore seems
unlikely that this emission peak is associated with a stationary zone
of warm gas at the base of the jet.  However, the same can not be said
for the [Fe~{\sc ii}] peak, for which we see no evidence for motion
downwind, even though radial velocity measurements indicate that the
emission is associated with higher jet velocities.  We postulate that
the [Fe~{\sc ii}] could be associated with a collimation shock at the
base of the jet.

\end{abstract}




\section{Introduction}

 

With high-spectral-resolution observations one may probe emission-line
regions at the bases of jets, and thereby hope to distinguish between
-- or at least constrain -- models of jet collimation and
acceleration. Indeed, there are a number of diagnostic atomic and
molecular lines in the near-infrared (near-IR) that are particularly
useful for studying the more deeply embedded Young Stellar Objects (YSOs)
\citep{dav01,dav03,pyo02,whe04,tak05,nis05}.  Most notable are the
[Fe~{\sc ii}] and H$_2$ lines, which are found to be bright at the base
of a number of jets from Class I protostars.  The emission is usually
blue-shifted, the [FeII] tracing a higher-velocity component than the
H$_2$, and offset along the axis of the jet by a few 10s of AU.  These
two species trace very different flow components: the [Fe~{\sc ii}]
derives from hot, dense, partially-ionised, gas (T$\sim$10,000\,K)
while the H$_2$ traces low-excitation, shocked (or possibly
fluoresced) molecular gas (T$\sim$2,000\,K).  These forbidden emission
line (FEL) and molecular hydrogen emission line (MHEL) regions exist
within a few hundred AU of each outflow source, and are therefore
either within or very close to the primary jet collimation and
acceleration zone.  The H$_2$ may be entrained along the walls of the
[Fe~{\sc ii}] jet, or it may trace a cooler molecular disk wind ejected
at larger disk radii.

As a further step toward a better understanding of these emission-line
regions, here we present adaptive-optics (AO) corrected H- and K-band
spectra of the MHEL and FEL regions associated with \object{SVS 13},
the driving source of the \object{HH 7-11} Herbig-Haro (HH) jet and
molecular outflow \citep{chr00,bach00}.  The source is located in the
L~1450 molecular cloud near the active low-mass star forming region
NGC\,1333 in Perseus. \object{SVS 13} is assumed to be at a distance
of 300~pc \citep{cer90,zee99}; modeling of the HH bow shocks in the
flow imply an inclination angle of $\sim$30\degr-50\degr\ with respect
to the plane of the sky \citep{har87,skd03}.  The {\em VLT}
observations presented here complement the earlier {\em U.K. Infrared
Telescope (UKIRT)} observations of
\citet{dav01,dav03}, which yielded important kinematic information
though lacked spatial resolution, and the more recent {\em Subaru}
observations of \citet{tak05}, where a thorough excitation
analysis has been made.

\section{Observations}


The combination of {\em NAOS} and {\em CONICA} at the {\em European
Southern Observatory (ESO) Very Large Telescope (VLT)} provides AO
correction using an IR-bright guide star in conjunction with near-IR
spectroscopy.  In our case, the jet source itself, \object{SVS 13}
(J$\sim$11.6, K$\sim$8.1), was used as the AO reference star. The
S27-3-SH and S27-3-SK modes used yield a wavelength
coverage of 1.37-1.72\micron\ and 2.02-2.53\micron\ at a spectral
resolution of R$\sim$1500. The slit width was 0.172\arcsec, the pixel
scale along the slit was 0.0270\arcsec ($\pm$0.0002\arcsec ) (measured
from the offsets of a bright star along the slit), and the slit length
was 28\arcsec.  The H- and K-band observations were conducted in
service mode on 13 and 17 January 2005, respectively.  With the slit
aligned along the \object{SVS 13} jet axis (position angle =
159$^{\circ}$ E of N), the telescope was nodded between object and
blank sky six times in each waveband.  Two 150\,sec exposures were
obtained at each position.

The individual spectral images were bad-pixel masked and flat-fielded
using observations of a halogen lamp.  Sky exposures were subtracted
from object frames before wavelength calibration using argon arc
spectra. {\em Starlink FIGARO} routines were used to correct for
distortion along the slit axis, so that the wavelength calibration was
constant along the spatial axis (along columns).  The continuum
associated with \object{SVS 13} was then registered along the slit
axis before co-addition of the data.  Individual sky-subtracted images
were compared to the shifted and co-added data in each waveband to
make sure that the spatial resolution had not been compromised by this
shifting and averaging process.  From Lorentzian fitting to profiles
taken through the \object{SVS 13} continuum at various wavelengths,
the full width at half maximum (FWHM) was found to vary between
0.24\arcsec\ and 0.40\arcsec\ in H, and 0.19\arcsec\ and 0.35\arcsec\
in K; in the averaged data the FWHM was 0.26\arcsec($\pm$0.03\arcsec)
in H and 0.23\arcsec($\pm$0.02\arcsec) in K.

The B5V bright star HIP 30943 (V=8.0) was observed as a telluric
standard.  However, the extracted H- and K-band spectra of this source
contain deep \hi\ brackett absorption lines.  Model Vega spectra were
therefore used to remove these lines.  Briefly, via an iterative
process similar to that described by \citet{vac03}, the model data
were scaled, smoothed and shifted before division into the HIP\,30943
spectra.  Comparison of the corrected HIP\,30943 spectra with
theoretical atmospheric transmission spectra presented by Vacca et
al. (2003) showed that this process worked very well; the small
difference in spectral type between Vega (A0V) and the standard (B5V)
had no noticeable affect on the \hi\ fitting.  The atmospheric
absorption features remaining in the corrected HIP\,30943 spectra were
subsequently aligned with equivalent features in the H- and K-band
\object{SVS 13} data, before the HIP\,30943 spectra were ``grown'' into
spectral images and divided into the \object{SVS 13} spectral images.





\section{Near-IR spectroscopy}

\begin{figure}
\epsscale{0.3}
\plotone
{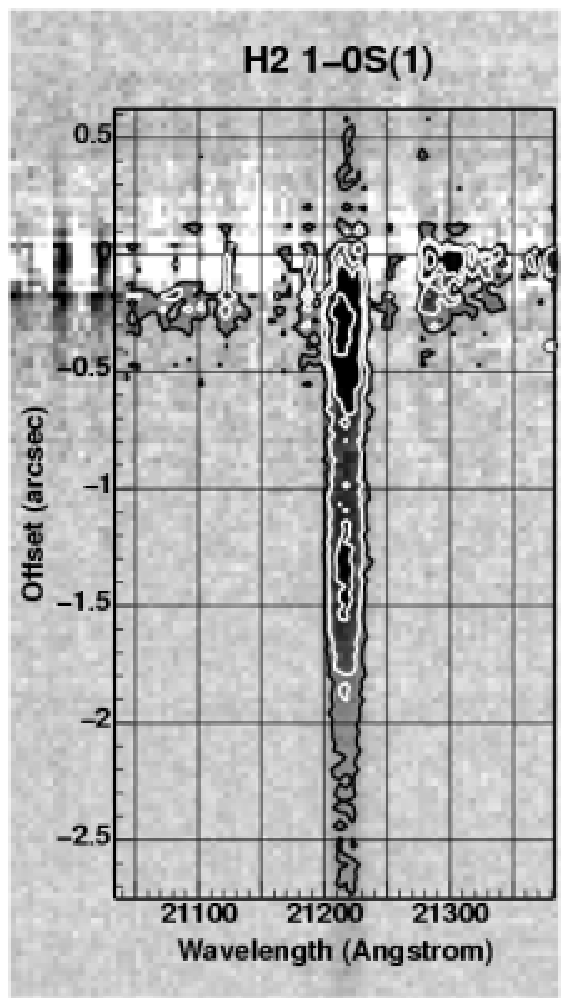}
\caption{Contour plot extracted from the K-band spectral image
showing the \htwo\ emission.  Negative offsets are along the south-eastern, 
blue-shifted jet axis.
The continuum associated with SVS\,13 has been 
fitted and removed (see text). 
\label{h2}}
\end{figure}

\begin{figure}
\epsscale{0.6}
\plotone
{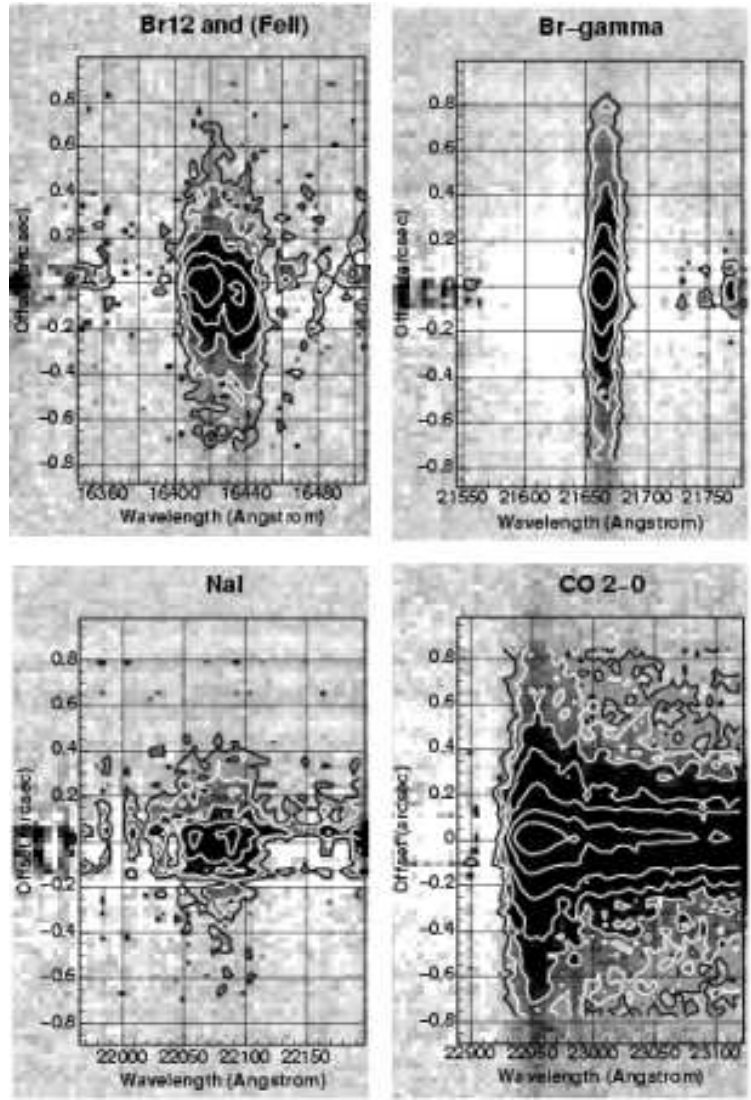}
\caption{Contour plots extracted from the H- and K-band spectral images
showing emission in select lines.  The continuum has  
again been removed.
\label{contours}}
\end{figure}

\begin{figure}
\epsscale{1.3}
\plotone
{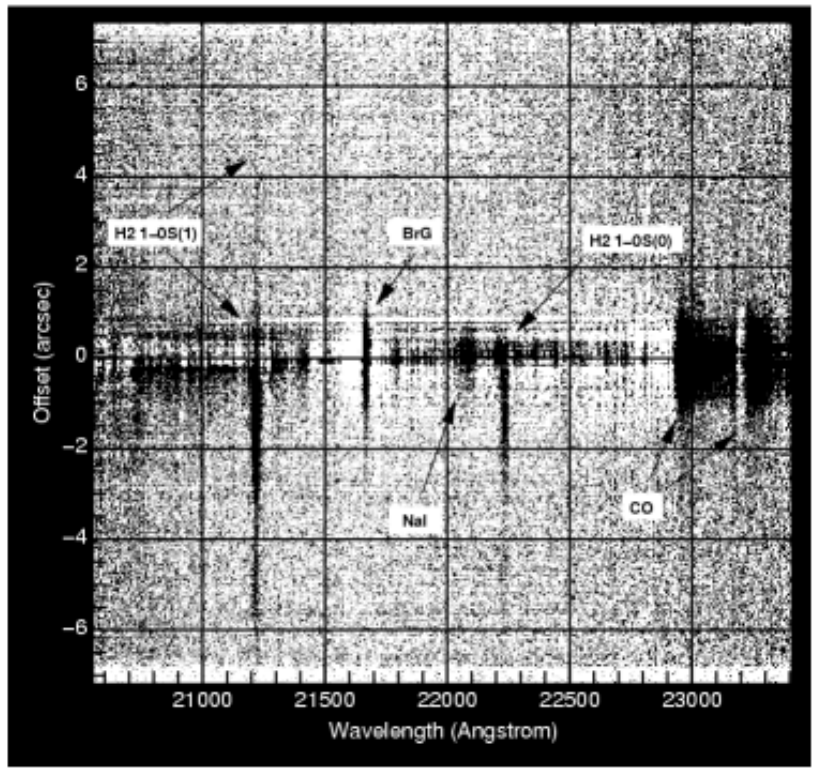}
\caption{Continuum-subtracted K-band spectral image with a grey-scale stretch 
set to show the faint H$_2$ emission along the slit. 
\label{faint}}
\end{figure}

H and K-band spectral images are shown in
Figures~\ref{h2}, \ref{contours} and \ref{faint}.  To more clearly show the
line-emission features, the \object{SVS 13} continuum has been removed from
each image, row by row, by fitting a second-order polynomial
to wavelength regions free of line emission.  However, residual shot
noise associated with the bright continuum does remain in some parts
of the data.

As is the case in other spectroscopic studies \citep{dav01,tak05},
extended emission is observed only in H$_2$.  However, for the first time,
faint H$_2$ emission is also observed in the \object{SVS 13} counter-jet
(positive offsets in Figure~\ref{faint}).  The H{\sc I} Brackett lines
in the H-band, and the CO, \brg\ and Na{\sc I}
lines in the K-band all appear as compact peaks coincident with the
continuum.  The [Fe {\sc ii}] peak is also compact, although it is slightly
offset along the blue-shifted jet.

In Figure~\ref{profiles} we compare plots of the \htwo\ and \fe\ profiles
traced along the slit, together with profiles of the adjacent
continuum.  The continuum plots are the average of profiles extracted
slightly blue-ward and red-ward of the line in each case.  The
continuum profiles effectively show the spatial resolution of the
observations, although they may be broadened slightly by nebulosity
associated with \object{SVS 13}.  Similar profile plots were produced
for the CO 2-0 bandhead and \brg\ line emission (not shown).

Only one component is evident in each of the CO, \brg\ and [Fe~{\sc ii}]
profiles, while as many as five separate peaks are identified in the
H$_2$ profile (labeled in Figure~\ref{profiles}).  From Lorentzian
fitting, the offsets with respect to the source continuum position and
the FWHM of the individual components seen in each line were measured.
These are given in Table~\ref{table1}.  The
CO and \brg\ peaks are unresolved and are precisely coincident, to
within a few AU, with the source continuum; we see no evidence for CO
or H{\sc i} components associated with the outflow. 

The [Fe~{\sc ii}] profile is, on the other hand, marginally resolved.
In the echelle data of \citet{dav03} and \citet{tak05} a radial
velocity of $\sim 140$~\kms\ with respect to the systemic velocity was
assigned to this component. Presumably this compact, high-velocity
emission derives from a discrete knot or shock front near the
base of the jet, rather than from an extended emission region.

In the H$_2$ data, we see an interesting trend where the size (FWHM)
of each component increases with distance from the source
(Table~\ref{table1}).  Moreover, if the H$_2$ and
[Fe~{\sc ii}] trace the same jet material, this trend extends to the
[Fe~{\sc ii}] data; the closer a component is to the source continuum,
the less extended it appears to be.  The H$_2$ components also
decrease in intensity with distance.  This behavior could be explained
by the cooling, decreased excitation and expansion of ``packets'' of
gas as they travel away from the source. A decrease in gas density and
excitation temperature is inferred from the excitation analysis of
\citet{tak05}.  A similar trend is also seen in small-scale jets from
T Tauri stars \citep{bac00,woi02}.  

Directly behind a fast-moving shock the very hot ($T \sim
10^3-10^5$~K) post shock gas will radiatively cool very rapidly,
before the region has time to expand. However, the bulk of the
material in the gas packet, {\em behind} the shock front working
surface, may expand adiabatically. If the pressure inside the warm ($T
\sim 100$~K) gas packet approaches that of the surrounding medium as it
expands, decreasing by a factor of $\sim$10-100, and if at the same
time the gas temperature in the packet decreases by a factor of 3-10
as it travels a distance of a few arcseconds (equivalent to the
inter-knot spacing), then the packet might be expected to increase in
volume by a factor of 1-30, or in diameter by a factor of
1-3. Isothermal expansion would result in greater expansion, while a
lower pressure gradient would suppress expansion.  Either way, the
ratio is potentially consistent with the relative sizes of the H$_2$
components in Table~\ref{table1}.


\section{Proper Motions}

By comparing the {\em VLT} data with the earlier {\em Subaru}
observations of \citet{tak05} we can measure, or at least set upper
limits on, the proper motions (PMs) of the [Fe~{\sc ii}] and H$_2$
components.  The {\em Subaru} observations, although at higher
spectral resolution ($R\sim1.1\times10^4$), were not obtained with AO
correction.  However, the {\em Subaru} pixel scale (0.060\arcsec
($\pm$0.002\arcsec )) does fully sample the seeing and the same slit
position angle (159\degr ) was used with both instruments.  Offsets of
emission features along the jet axis can therefore be measured
accurately and a direct comparison made between the {\em VLT} and {\em
Subaru} observations (see Table~\ref{table2}).  Note, however, that
the {\em Subaru} slit was slightly wider than the {\em VLT} slit;
0.30\arcsec\ versus 0.172\arcsec.

\begin{figure}
\epsscale{0.5}
\plotone
{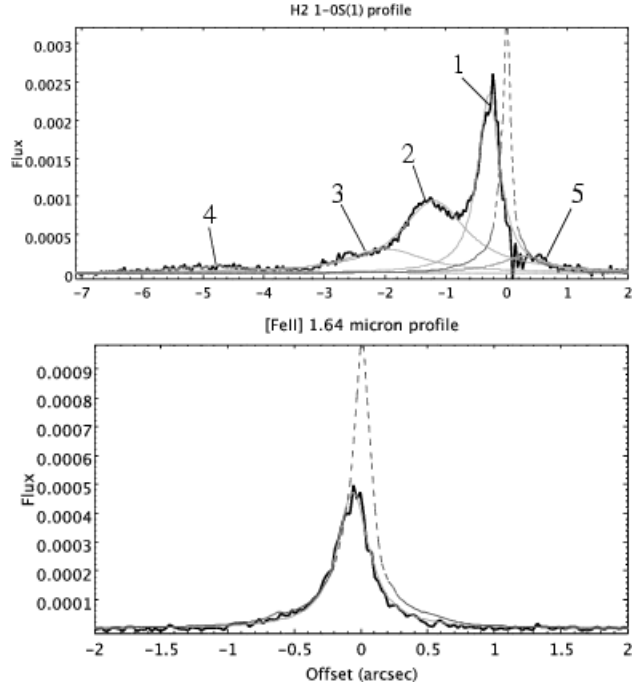}
\caption{Profile plots showing the distribution of \htwo\ and \fe\ 
emission along the slit.  The fine grey lines show Lorentzian fits to
the data; one fit for the [Fe~{\sc ii}] peak though five fits to the
H$_2$ data.  The dashed line shows the profile through the continuum
adjacent to each line. \label{profiles}}
\end{figure}
\begin{figure}
\epsscale{0.5}
\plotone
{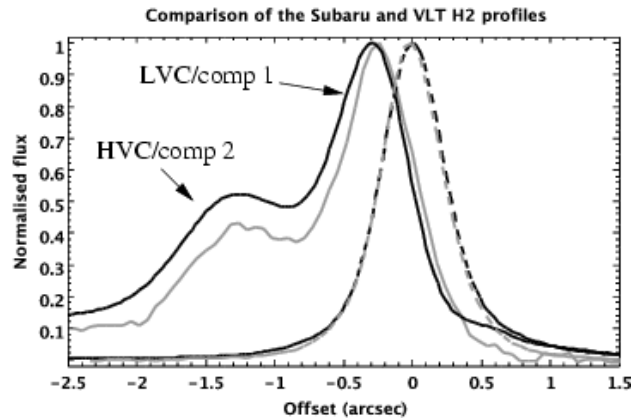}
\caption{Comparison of the {\em Subaru} (Grey) and {\em VLT} (black) H$_2$
profiles.  The dashed lines show the adjacent continuum data.  The VLT
observations have been Gaussian-smoothed to the spatial resolution of
the Subaru data. \label{profiles2}}
\end{figure}

\begin{itemize}

\item In the {\em VLT} H-band observations, the single [Fe~{\sc ii}] component
in Figure~\ref{profiles} may correspond to the high-velocity peak seen
in the {\em Subaru} data, for which \citet{tak05} assign a radial
velocity of $\sim 140$\kms .  Gaussian fitting yields an offset of
0.10\arcsec ($\pm$0.01\arcsec ) for the {\em Subaru} peak.  Curiously,
the offset of the {\em VLT} peak in Table~\ref{table2} implies that
this feature has moved upwind, i.e. {\em closer} to the source.  We
certainly see no evidence for pronounced movement away from the jet
source.

\item In the {\em Subaru} K-band echelle data two velocity components 
are resolved in H$_2$; a bright, low-velocity component (LVC)
blue-shifted by $\sim$30\kms\ that is spatially offset along the jet
by 0.23\arcsec ($\pm$0.01\arcsec ), and a fainter, more diffuse,
high-velocity component (HVC) blue-shifted by $\sim$100\kms\ that is
offset by 1.27\arcsec ($\pm$0.02\arcsec ).  These two peaks, which are
spatially as well as kinematically separate, probably correspond to
the first two components in Figure~\ref{profiles}.  In
Figure~\ref{profiles2} we compare the {\em Subaru} data with smoothed
{\em VLT} data: here the components clearly have a similar spatial
offset, width and brightness ratio.  (Note that if the data were
normalized to the HVC/component 2 peak, then the {\em Subaru} LVC
would be brighter than the {\em VLT} component 1.  Given the wider
{\em Subaru} slit, one would expect this to be the case if the LVC
traced a broader flow component than the HVC.) For the LVC/component
1, the offsets in Table~\ref{table2} indicate a shift downwind of
0.06\arcsec ($\pm 0.02$\arcsec ) over the 783~day time interval
between the {\em Subaru} and {\em VLT} observations.  However, the
HVC/component 2, like the high-velocity [Fe {\sc ii}] component, again
exhibits no measurable PM away from the source.

\end{itemize}

From the radial velocities of the HVC and LVC, and assuming an inclination 
angle of 40\degr\ to the line of sight and a distance of 300\,pc to
\object{SVS 13}, then over the time period between the observations the 
H$_2$ LVC and HVC should have moved by 0.04\arcsec\ (12~AU) and
0.13\arcsec\ (38~AU) on the sky, while the [Fe~{\sc ii}] feature should
have moved by 0.18\arcsec\ (53~AU).  Clearly, only the H$_2$ LVC has a
PM consistent with the observed radial velocity.  Such consistency is
expected if, for example, the gas is accelerated by a fast-moving
shock front in a heavy jet.

The apparent lack of movement of the H$_2$ HVC could result from: (1)
the larger errors associated with the positional measurements of this
fainter, more extended feature; (2) blending of component 2 with
component 3 (Table~\ref{table1}) in the lower-spatial-resolution {\em
Subaru} data, so that the overall peak appears shifted downwind;
or (3) the fact that the time interval between the observations is
comparable to the molecular cooling time and therefore the timescale
for morphological change, which could introduce unknown errors on the
PM measurements.

The lack of a measurable PM in the [Fe {\sc ii}] peak is more difficult
to explain.  If the [Fe {\sc ii}] peak was associated with ejected
clumps or bullets, then one would perhaps expect to see additional
components along the flow axis (as is the case in H$_2$).  Instead,
the [Fe~{\sc ii}] is confined to a single, very compact peak at the jet
base.  The [Fe~{\sc ii}] could be associated with a
stationary, collimating shock, similar to that described in the models
of \citet{ouy94}.  If this is indeed the case, the collimation point
would be at a distance of $\sim 20$\,AU from the source.  

On the other hand, we may be observing two completely independent
[Fe~{\sc ii}] peaks in the {\em Subaru} and {\em VLT} data, rather than
the same near-stationary feature.  The {\em Subaru} feature may have
faded, to be replaced by a new peak in the {\em VLT} observations.
The cooling time, from 20,000~K to 7,000~K for dense
($10^{5}-10^{6}$~cm$^{-3}$), post-shock gas will be of the order of
weeks or even days \citep{smi03}, so morphological changes are
certainly possible.  This might explain the apparent ``upwind''
movement of this feature.

A number of groups have attempted to {\em image} the emission-line
region at the base of the SVS\,13 outflow.  \citet{dav02} used a
Fabry-Perot (FP) etalon to ``boost'' the line/continuum ratio in their
data.  The H$_2$ emission at the jet base is certainly evident in
their image, although the poor spatial resolution of their
observations ($\sim$1\arcsec) somewhat limits their ability to extract
spatial information.  They do collapse their image along an axis
perpendicular to the jet and attempt to subtract the continuum
emission from the resulting profile (their Figure 4).  They identify a
``peak'' and a ``plateau'' along the jet axis which may correspond to
H$_2$ components 2 and 3 in the {\em VLT} data.  We list the offsets
of these features in Table~\ref{table2}.

\citet{nor02} used the {\em HST} to obtain high-spatial resolution
near-IR images of the HH\,7-11 outflow.  They present a
continuum-subtracted H$_2$ image of SVS\,13 (their Figure 3).
However, residual artifacts left over from the continuum subtraction,
in the bright core but also in a ring of radius $\sim$1\arcsec, make
identifying features within 0.5--1.0\arcsec\ of SVS\,13 more-or-less
impossible.  However, they do detect and resolve the H$_2$ emission in
the flow at offsets of $>$1\arcsec . Within 1\arcsec-2\arcsec\ of
SVS\,13 the H$_2$ appears to comprise at least two knots superimposed
on to a patch of moderately extended emission, suggesting a broad
opening angle for the H$_2$ flow.  Component 2 in the {\em VLT} data
(the HVC in the {\em Subaru} observations) may therefore constitutes
{\em two knots}; we give a combined offset for these two features
(measured along the {\em VLT/Subaru} slit axis) in Table~\ref{table2}
for reference.

Overall, the results from these two imaging surveys can only be used
to confirm the presence of emission features.  Uncertainties in the
continuum subtraction limit their usefulness when measuring PMs.
Notably, the compact LVC evident in both the {\em Subaru} and {\em
VLT} spectroscopy was not extracted from the FP or {\em HST} images.

Clearly, follow-up observations with the {\em NACO} system are needed
to better constrain the PM measurements discussed above, particularly
for the [Fe {\sc ii}] peak and the H$_2$ components further downwind.
Confirmation of the ``stationary'' [Fe~{\sc ii}] peak would add
credence to the collimation-shock interpretation.  If we also
ultimately find that the H$_2$ peaks do {\em not} move, then in H$_2$
we could be witnessing a warm, stationary region or perhaps a
turbulent mixing layer between the jet and ambient medium at the jet
base through which jet material flows, rather than discrete bullets or
clumps ejected during episodes of increased accretion.  However, given
the tentative result for the LVC, and the fact that knots in
Herbig-Haro jets and molecular flows typically do have large proper
motions \citep{rei01}, for H$_2$ at least, this seems unlikely.

Finally, we mention that the dynamical age of the H$_2$ LVC, components 1
in the {\em VLT} data, when derived from its PM and offset in
Table~\ref{table1}, is $\sim 10$\,yrs.  Notably, the optical outburst
reported by \citet{eis91} occurred roughly 15\,yrs before the
observations reported here.  Our results are thus reasonably
consistent with the H$_2$ LVC being produced by this outburst and the
associated increase in accretion.




\clearpage




\clearpage

\begin{table}
\begin{center}
\caption{Parameters from the profile fitting\label{table1}}
\begin{tabular}{lrrr}
\tableline\tableline
Line\tablenotemark{a}    & Offset      & Offset    &    FWHM  \\
                         & (arcsec)    & (AU)      & (arcsec)             \\
\tableline
H$_2$ - 1          & 0.29($\pm$0.009)  & 87.8($\pm$3)     & 0.44 \\
H$_2$ - 2          & 1.19($\pm$0.018)  & 358($\pm$6)      & 1.3  \\
H$_2$ - 3          & 2.0($\pm$0.1)     & 600($\pm$30)     & 1.5  \\
H$_2$ - 4          & 4.8($\pm$0.2)     & 1428($\pm$30)    & 1.8  \\
H$_2$ - 5\tablenotemark{b} & 0.28($\pm$0.1) & 86($\pm$30) & 1.0  \\
${\rm [FeII]}$     & 0.061($\pm$0.004) & 18.3($\pm$1.2)   & 0.29 \\
\brg               & 0.001($\pm$0.003) & $<$0.9           & 0.19 \\
CO                 & 0.004($\pm$0.003) & $<$1.2           & 0.17 \\
\tableline
\end{tabular}
\end{center}
$^{a}${Lorentzian fits to five components in H$_2$ (see
Figure~\ref{profiles}); a single Lorentzian is fitted to the single peak in 
each of the extracted CO, [Fe~{\sc ii}] and \brg\ profiles.} \\
$^{b}${Possible component in the red-shifted counterflow (positive offsets in
Figure~\ref{profiles}).}
\end{table}
                                                                                
\begin{table}
\begin{center}
\caption{Component offsets and proper motions\label{table2}}
\begin{tabular}{lccccc}
\tableline\tableline
Line/component\tablenotemark{a} & {\em VLT}\tablenotemark{b}    & {\em Subaru}\tablenotemark{c} & {\em UKIRT}\tablenotemark{d} 
    & {\em HST}\tablenotemark{e}         & P.M.\tablenotemark{f}           \\
                         & Offset (\arcsec) & Offset (\arcsec) &  Offset (\arcsec) & Offset (\arcsec) & (\arcsec /yr)  \\
\tableline
H$_2$ - 1      & 0.29($\pm$0.009)  & 0.23($\pm$0.01) & ...             &   ...                                & 0.028($\pm$0.007) \\
H$_2$ - 2      & 1.19($\pm$0.018)  & 1.27($\pm$0.02) & 1.28($\pm$0.01) & $\sim$1.1($\pm$0.5)\tablenotemark{g} & $<$0.05 \\
H$_2$ - 3      & 2.0($\pm$0.1)     &  --             & 2.13($\pm$0.09) &   ...                                & $<$0.03 \\
H$_2$ - 4      & 4.8($\pm$0.2)     &  --             & ...             &   ...                                & ...     \\
${\rm [Fe II]}$ & 0.061($\pm$0.004)& 0.10($\pm$0.01) & ...             &   ...                                & -0.018($\pm$0.006) \\ 
\tableline
\end{tabular}
\end{center}
$^{a}${Same as Table~\ref{table1}}.\\
$^{b}${{\em VLT} spectroscopy presented in this paper; data obtained on 1/17/2005.} \\
$^{c}${{\em Subaru} spectroscopy presented in  \citet{tak05}; data obtained on 11/25/2002.} \\
$^{d}${{\em UKIRT} Fabry-Perot imaging from \citet{dav02}; data obtained on 11/23/2000.} \\
$^{e}${{\em HST-NICMOS} imaging from \citet{nor02}; data obtained on 1/9/1998.} \\
$^{f}${Proper motion measurements or upper limits; the negative value for
the \fe\ peak reflects its apparent motion {\em towards} the source.} \\
$^{g}${Multiple components spread over an area of roughly 1\arcsec .}
\end{table}
                                                                                
 

\end{document}